# Simple and open 4f Koehler transmitted illumination system for low-cost microscopic imaging and teaching

Jorge Madrid-Wolff[1], Manu Forero-Shelton[2]

1- Department of Biomedical Engineering, Universidad de los Andes, Bogota, Colombia

2- Department of Physics, Universidad de los Andes, Bogota, Colombia

anforero@uniandes.edu.co

**ORCID:**

JMW: https://orcid.org/0000-0003-3945-538X

MFS: https://orcid.org/0000-0002-7989-0311

**Any potential competing interests:** NO

**Funding information:** 1) Department of Physics, Universidad de los Andes, Colombia, 2) Colciencias grant 712 "Convocatoria Para Proyectos De Investigación En Ciencias Básicas"  3) Project termination grant from the Faculty of Sciences, Universidad de los Andes, Colombia.

**Author contributions:**

JMW Investigation, Visualization, Writing (Original Draft Preparation)

MFS Conceptualization, Funding Acquisition, Methodology, Supervision, Writing(Original Draft Preparation)

# Title

Simple and open 4f Koehler transmitted illumination system for low-cost microscopic imaging and teaching

# Abstract

Koehler transillumination is a powerful imaging method, yet commercial Koehler condensers are difficult to integrate into tabletop systems and make learning the concepts of Koehler illumination difficult. We propose a simple 4f Koehler illumination system that offers advantages with respect to building simplicity, cost and compatibility with tabletop systems, which can be integrated with open source Light Sheet Fluorescence Microscopes (LSFMs). With those applications in mind as well as teaching, we provide instructions and parts lists to assemble the system. The system´s performance is comparable to that of commercial condensers and significantly better than LED flashlights such as those found in low-cost diagnostic devices and LSFMs with respect to illumination homogeneity and sectioning capability.

# Introduction

Transmitted illumination is one of the most commonly used light microscopy techniques, and it is also used as a complementary method to find samples in other kinds of microscopy such as fluorescence microscopy thanks to its gentleness with the samples as well as the complementary information it provides. August Koehler's introduction of his illumination system in the late 1800's radically improved the attainable quality of microscopic imagery (1). Before Koehler's system, samples were unevenly illuminated, both spatially and angularly, typically under critical illumination, in which the lamp is placed at the back focal plane of a condenser lens. This resulted in conjugate images of the lamp on the sample, a strong spatial dependence of resolution, the projection of shadows, heating due to imaging the light source on the sample plane, and glare from the illumination source in the detection objective affecting image contrast. Koehler illumination significantly improves illumination quality, in several aspects including homogeneity, glare, and resiliency of the illumination homogeneity with respect to dust or imperfections on the condenser lens. Moreover, it provides control of the illumination´s angular profile via changes in the condenser´s numerical aperture (NA). Additional advantages of an angularly and spatially homogeneous illumination include adjusting contrast and depth of field (2). These advantages are highly beneficial, even in systems that are not mainly dedicated to transmitted illumination imaging such as Light Sheet Fluorescence Microscopy (LSFM). For instance, when imaging complex 3D samples, that have features at different depths, the possibility of adjusting contrast and depth of field via changes in the NA may facilitate navigating through the sample or acquiring reference tomographic views. From the perspective of low-cost microscopy, Koehler illumination provides several advantages. Homogeneous illumination makes it easier to analyze samples, especially for automated analysis (3,4) in the context of low-cost diagnostics (5). Additionally, in difficult environments, such as those found outside research and diagnostic laboratories, the effect of dust or imperfections in the condenser lens are minimized in Koehler illumination, reducing image degradation over time.

Commercial Koehler illumination condensers have drawbacks for teaching, tabletop systems, and low-cost microscopy. From the teaching perspective, in our experience and that of others (6), Koehler illumination is one of the teaching challenges in introductory microscopy courses. Commercial condensers make it hard for students to grasp the functioning of the system, due to the optical complexity required to make them compact and optically optimized, as well as their closed-box nature. From the perspective of microscopes built on optical tables, they are hard to integrate, and they are relatively costly since they are not fundamental for the core function of the microscope. For example, in our experience, few labs that do LSFM use proper Koehler illumination for transmitted light viewing, as most data comes from fluorescence despite this making it it harder to correlate anatomy in samples with fluorescence. There are examples in LSFM where transmitted light is used advantageously, such as Optical Tomography (7) and Optical Projection Tomography (8,9), but the first requires another water immersion objective, making it harder to implement, and the second is not commonly used in non-specialist labs. Finally, in the area of low-cost microscopy, many systems work with a flashlight to illuminate the sample (4,10–12), limiting the potential of those systems in terms of image quality.

Here we show a simple implementation of a 4f Koehler illumination setup that can be made to be low-cost for applications in both teaching, low cost diagnostics and LSFM. The 4f implementation makes alignment as well as understanding of the system easy, and we include a parts list as well as building and alignment instructions. This system is highly intuitive, illustrating the need for even illumination while simplifying its design and operation. The system can be built, aligned and used by non-experienced users.

## Materials and Methods

### Koehler illumination condenser and light source

The setup consists of a light emitting diode, whose light we collect and collimate with an aspheric lens. Two convex lenses in a 4f configuration relay the focal plane of the aspheric onto the focal plane of the detection objective of the microscope, where the sample is placed. Irises at the common focal planes of the lenses act as field and aperture diaphragms.

In the simplest implementation, the light source is a mounted white LED (Thorlabs, MWWHL4) with its corresponding controller (Thorlabs, LEDD1B) and power supply (Thorlabs, KPS101). Light from the LED is collected with an aspheric lens with a diffusing surface in order to minimize system size (Thorlabs, ACL2520U-DG6). The front focal plane of the aspheric lens is relayed by means of two achromatic lenses (Edmund Optics, 47-637) onto the sample plane. Cage-system compatible irises (Thorlabs, CP20S) at the common focal planes of the lenses, work as field and condenser aperture diaphragms as shown in figure 1. Although not indispensable, we use a right- angle kinematic mirror mount (Thorlabs, KCB1/M) between the two lenses to make the system more compact in the 4f setup. The NA of the condenser is given by the focal length of the last lens, which in this case is of 50mm. Hence the NA is n*sin(arctan(10mm/50mm)) = 0.2, given that the maximum aperture of the condenser lens is 20mm (radius 10mm).

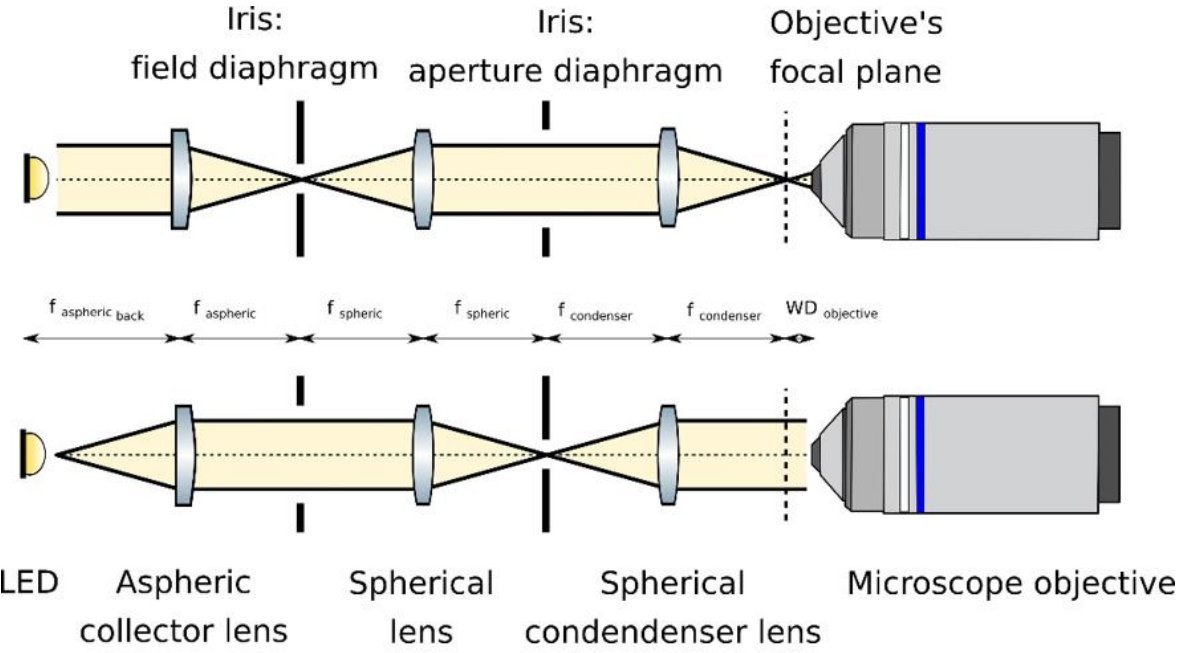

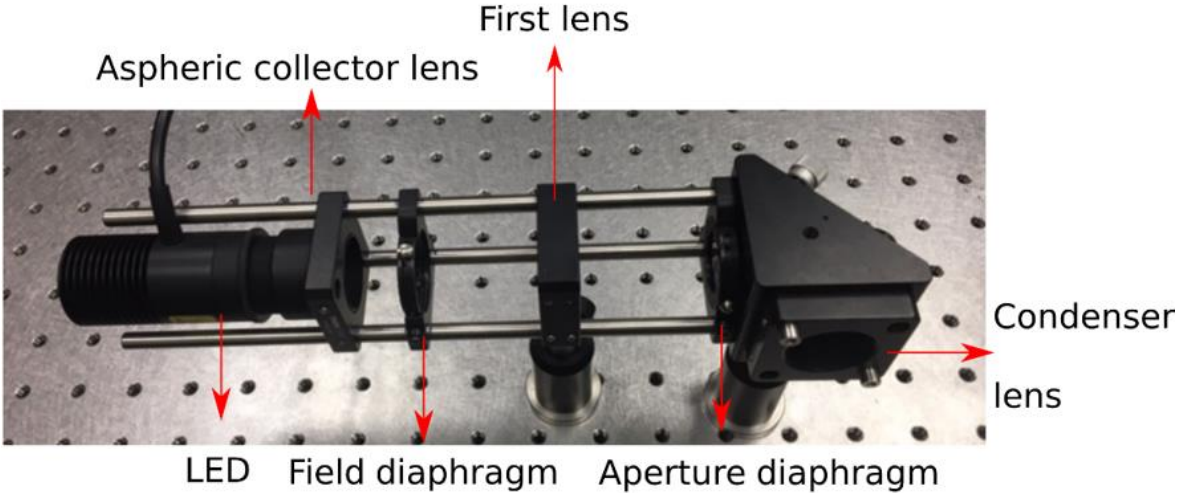

Fig 1. **Optical setup.** (A) Schematic of the optical setup. The system is all configured in 4f, conjugating two sets of planes. The upper section of the diagram illustrates the conjugation of the sample plane with the field diaphragm and (not shown in the schematic) with the image plane on the camera sensor. In these planes the image of the LED is lost after having been Fourier transformed by the collector lens and spatially homogenized by the diffusing surface. (B) Photograph of the setup (right-angle mirror is optional).

The illumination system is built as a separate unit from the microscope, as described in supplementary note S1-Assembly Instructions. Following the procedure in the instructions ensures that the field and aperture irises are at their correct relative positions. To place the system at the correct position with respect to the optical train of the microscope´s detection path it is necessary to close the field diaphragm and move the system along the optical axis until an image of the diaphragm appears at the camera sensor (or the eyepieces). This can either be done manually, or with the help of one or up to three translation stages. In our experience a single translation stage (in the direction of the optical path) or an XY stage is enough to simplify the alignment procedure with the microscope in the long term, since the height of the lenses on a tabletop horizontal microscope varies little. Furthermore, final adjustments to the position of the LED may be made with help of the microscope´s detection. With the system in place, using a digital camera in live view it is possible to obtain a profile of the intensity across the field. This live profile can be viewed using µManager (13) and ImageJ (14) in conjunction as per the instructions. The position of the LED diode is finally adjusted by rotating the lens tube until the flattest intensity profile possible is achieved.

## Comparative systems and measurements

Field homogeneity was measured using a scientific CMOS camera (Hamamatsu, ORCA Flash 4.0 v2) mounted directly at the output of the condenser, as lower cost cameras we tested had noise that was larger than the expected homogeneity. We made sure that the camera's chip was placed at the focal plane of the condenser by first acquiring a sharp image of the field diaphragm. We inserted a neutral density filter close to the aperture diaphragm to reduce the light intensity and avoid saturation. The pixels were binned 4x4 in the image post-processing to limit the effect of camera noise on the local homogeneity measurements.

A Nikon Ti Eclipse inverted microscope was used to compare the 4f setup with a commercial system. The condenser has a maximum NA of 0.52 and a working distance of 30mm (Nikon TI-C-LWD). The Koehler illumination system of the commercial microscope was aligned by adjusting the $x, y$, and $z$ positions of the field diaphragm across and along the optical axis.

Finally, a generic LED flashlight bought at a local hardware store (Sky Light rechargeable LED torch), with no additional optics, which is a commonly used illumination system in low-cost microscopy as well as LSFM was used to compare the 4f system with the other end of the performance spectrum.

## List of parts

| Supplier | Part | Part # | Quantity | Price [USD] | Sub total [USD] | Notes |
|---|---|---|---|---|---|---|
| | **LED illumination** | | | | | |
| Thorlabs | White LED | MWWHL4 | 1 | 172,38 | 172,38 | Illumination source |
| Thorlabs | LED controller | LEDD1B | 1 | 304,98 | 304,98 | |
| Thorlabs | Power supply for LED controller | KPS101 | 1 | 33,33 | 33,33 | |

| Thorlabs | SM1 Lens Tube | SM1L03 | 1 | 12,16 | 12,16 | Simplifies collimation with the condenser lens |
|---|---|---|---|---|---|---|
| Thorlabs | Adjustable Lens Tube | SM1V05 | 1 | 30,25 | 30,25 | |
| Thorlabs | Aspheric Condenser Lens w/ Diffuser, Ø25 mm, f=20.1 mm, NA=0.60, 600 Grit | ACL2520U-DG6 | 1 | 18,46 | 18,46 | Light collector. Make sure the round face faces away from the LED |
| | **LED llumination total** | **571,56** | | | | |
| | | | | | | |
| | **Koehler Optics** | | | | | |
| Thorlabs | Cage plate (lens mount) | CP02/M | 2 | 16,4 | 32,8 | Mount the koehler lenses |
| Thorlabs | Cage plate (lens mount) Thick | CP02T/M | 1 | 21,93 | 21,93 | Thick Mount for using a single support for the system |
| Thorlabs | Cage system iris | CP20S | 2 | 88,49 | 176,98 | Condenser and field diaphragms |
| Thorlabs | Cage rods 18" | ER18 | 3 | 25,5 | 76,5 | |
| Thorlabs | Cage rods 1" | ER1-P4 | 1 | 19,19 | 19,19 | Align condenser lens |
| Edmund Optics | 50mm FL, Achromatic Lens | 47-637 | 2 | 91,5 | 183 | 4f achromatic lenses, one is the condenser lens, that could be changed for longer or shorter focal lengths depending on application and desired maximum N.A. |
| | | | | | | |
| | **Support** | | | | | |
| Thorlabs | Optical post | TR50/M | 1 | 5,19 | 5,19 | |
| Thorlabs | Optical post holder | PH40E/M | 1 | 23,97 | 23,97 | These magnetic post holders make alignment easier on a steel surface if you don´t use a translation stage . |
| | | | | | | |
| | **Total without optional parts** | **1111,12** | | | | |

**Table1. List of parts of the 4f Koehler transmitted light illuminator**. List of the parts required for implementing the easiest version of the illumination condenser.

The list above is optimized for simplicity, not cost, despite being cheaper than many commercial condensers. For a lower-cost alternative see supplementary table 1. The main changes have to do with the use of a cheaper diode as a light source, and the final cost is under half of this system´s. Optional parts for making the system more compact and simplifying the translation can be found in supplementary table 2.

## Results

### Field homogeneity is better than for led flashlights

We evaluated the homogeneity of the illumination by placing a camera at the focal plane of the condenser lens in order to eliminate the effects from the apodization of the detection objective. We placed a neutral density filter in the system and reduced the camera's exposure time to 30ms to avoid saturation. Figure 2 compares the homogeneity of the illumination over a field of 1300µm by 1300µm, which corresponds to the field of view achieved with a large chip scientific CMOS camera and a 10x objective. The 4f system achieves a more homogeneous field, with differences in intensity of up to 5%, compared to ~20% for the LED flashlight.

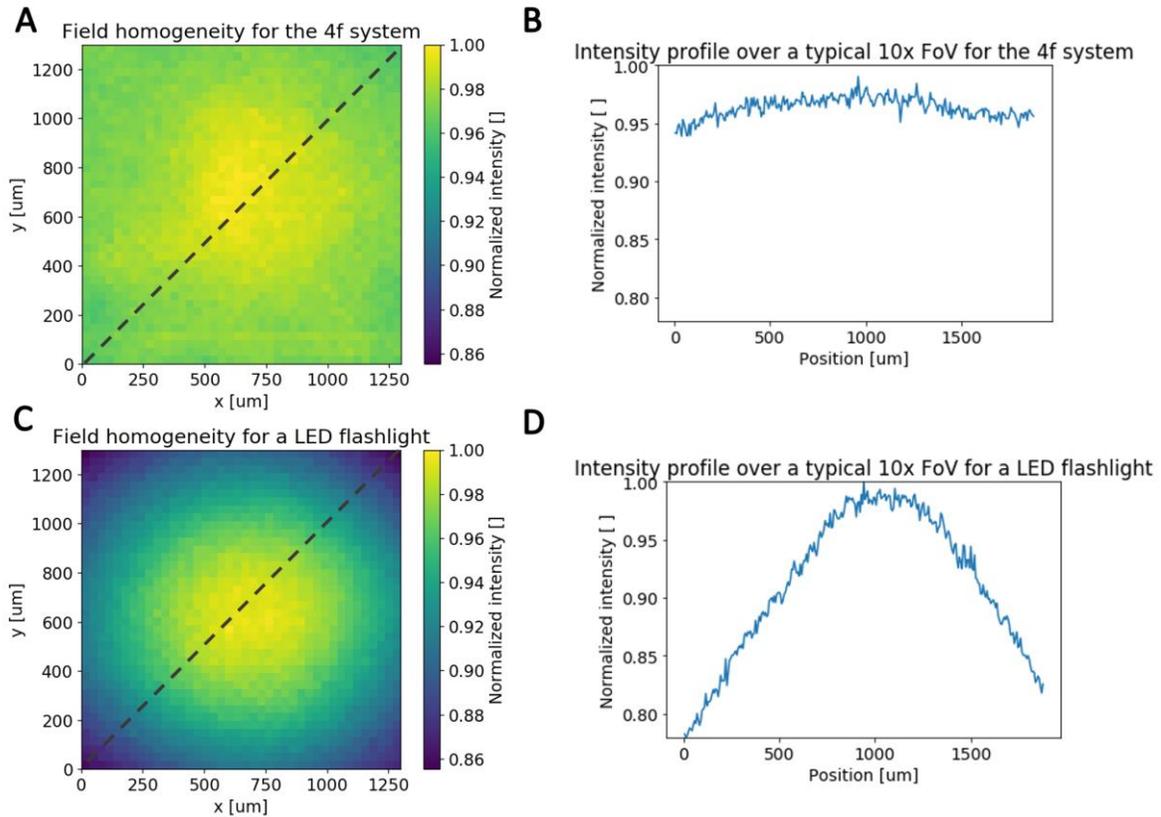

Fig 2. **Field homogeneity.** (A) Relative intensity profile for the system presented here over a typical field of view with a 10x magnification. (B) Intensity profile along the dotted diagonal in (A). (C) and (D), analogous to (A) and (B) but for a single LED flashlight.

Sectioning capability is superior to that of led flashlights and comparable to that of commercial systems.

In many cases, high numerical-aperture illumination is desirable, as it reduces contrast and shortens the depth of field, which is useful when imaging samples with fine features at different depths. We compared the 4f system against a commercial one, both at their highest NA (0.2 and 0.52 respectively) and with an LED flashlight (whose NA is not adjustable). We imaged the petals of an orchid (*Phalaenopsis amabilis)* at 10x magnification with a detection NA of 0.3 at different depths, as shown in figure 3. Under these conditions, it is possible to distinguish features at some planes which are not visible in other planes, e.g. the nuclei and fibers that are observable at z=0um and z=20um are not observable in the other planes, when illuminating with the 4f system and the commercial one. When illuminating with the LED flashlight, all features appear overlaid, making it impossible to distinguish specific structures such as nuclei and cell walls, among others. This system achieves an NA comparable, although not as high, as that of the commercial system, which is evidenced in the higher contrast and longer depth of field of the views in the top row of figure 3.

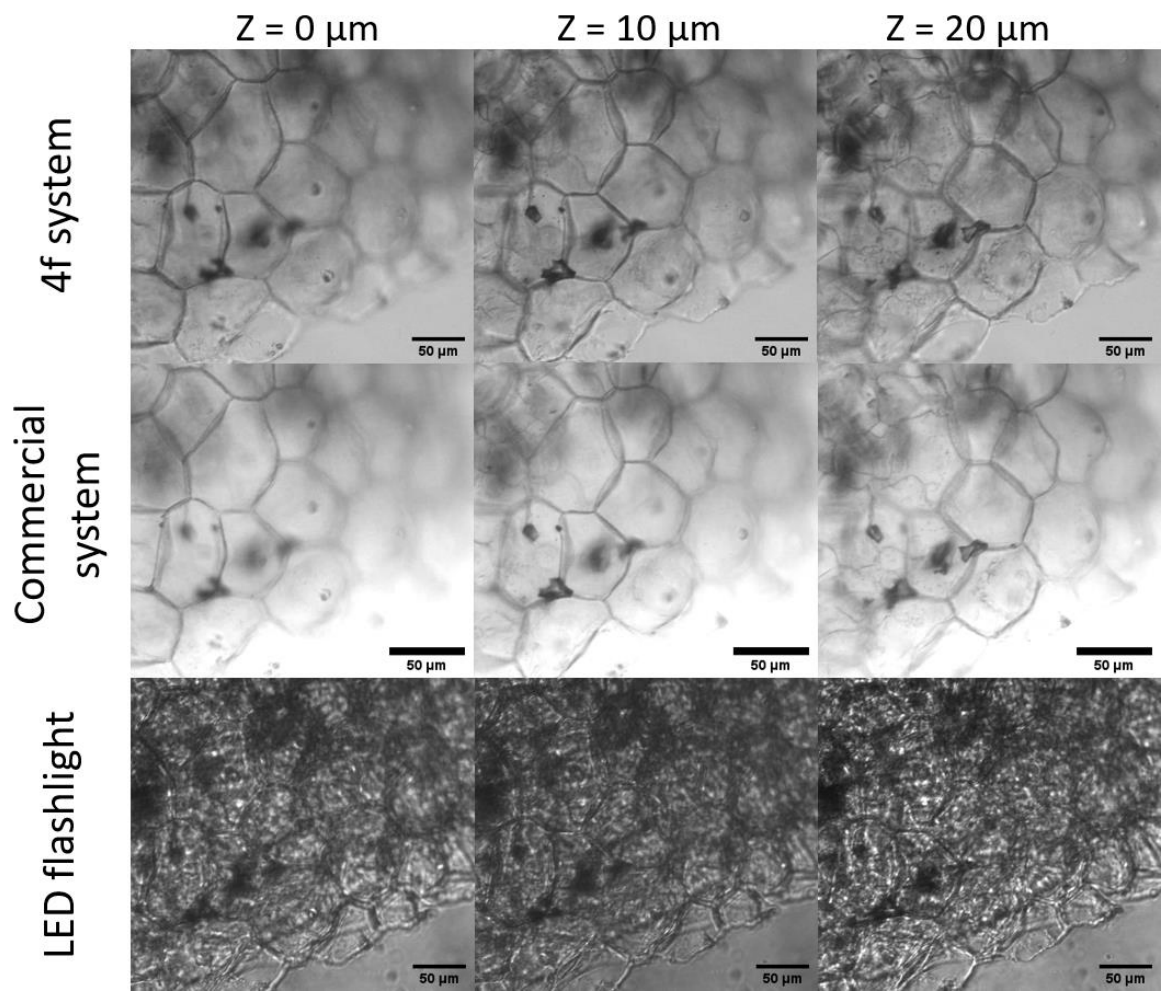

Fig 3. **Sectioning capability of the 4f system with respect to a commercial condenser and an LED flashlight.** Different planes of a petal from *Phalaenopsis amabilis*, captured with a Nikon 10X 0.3

air objective. (*Top row*) 4f system with the condenser aperture fully open (NA=0.2), (*middle row*) Nikon Ti Koehler system with the condenser aperture fully open (NA=0.52), and (*bottom*) Single LED source. Scale bar 50um.

## Discussion-conclusion

We presented a simple to assemble 4f Koehler illumination system for transmitted-light imaging in microscopy. We examined the spatial homogeneity of the illumination and showed that over regions of 1300µm by 1300µm, which corresponds to the field of view of a large sensor scientific CMOS camera with a 10x objective, illumination intensity is fairly homogeneous. Additionally, we compared its sectioning ability against a commercial system and an LED flashlight for low NAs, and found that image quality was much better than that of the LED flashlight and comparable to that of a commercial system.

There are several variations of this setup that could be implemented, depending on the available budget and experimental requirements. This simple-to-implement version of the illuminator costs around $1500 USD, where it starts being comparable in optical quality, and still cheaper than a commercial standalone condenser, whose cost including a state-of-the-art LED could be up to $7000 USD (See supporting table 3 for details). On the other end of the spectrum it should be possible to build the system for under $500 USD (Supporting table 1), although at the expense of a more complicated assembly due to the work required to assemble the electronics of the illumination. Other simple variations of the system include changing the condenser lens for a shorter focal length for better NA, at the expense of a shorter working distance than the current 50mm and of the turning mirror.

We provide assembly and alignment instructions for the 4f system (Supporting note 1), which can be built and put in place by someone with little experience with tabletop optics. We do not include instructions for alignment as there are many resources online covering that topic. Additionally, this system illustrates the inner functioning of Koehler illumination systems, which generally operate as a closed box and are obscure to many microscope users. The 4f configuration makes it easier to align and understand the concept of conjugate planes, and how the illumination is controlled in the Koehler configuration, although it is less compact than an equivalent system without 4f alignment.

Since this system is compatible with tabletop instruments it can be easily implemented in laboratories working with custom microscopes, such as light sheet systems. Specifically, this illumination system is compatible with open-source options for LSFM (15,16), complementing the description for a low-cost critical illumination system (17). This illumination setup can enhance image quality in complementary transmitted-light imagery, sample navigation, provide anatomic context for fluorescent structures (7) and could also improve quantitative fluorescence by reducing the effect of attenuation artifacts in LSFM (9).

A key application of image acquisition in microscopy is sample analysis. In many cases, researchers or clinicians using the micrographs expect quantitative metrics from numerous samples. To achieve this, automated image analysis is conducted on the acquired data. The reproducibility and confidence of automated analysis strongly depends on the quality of the imagery that it receives as input. Additional post-capture processing can be performed on the imagery to enhance its quality; however sometimes post-processing is not enough to compensate for limited quality in the recorded data (18). Non-uniform illumination, saturated regions, fluctuating brightness, and excessive cluttering are examples of factors limiting the quality of acquired images (19,20) and the reproducibility of quantitative analyses made from them (21).

The open architecture of this system should help the spread of improved illumination in low-cost diagnostics and microscopy.

## Acknowledgements

We thank Jhonny Turizo and his assistant Jhon Castillo for the design of the low-cost LED driver system as well as the holder, with additional help from Luis Gómez. The Department of Physics and Colciencias grant 712 "Convocatoria Para Proyectos De Investigación En

Ciencias Básicas"  contract 112-2016 financed materials for the project, with an additional contribution from the "project termination" grant from the Faculty of Sciences. We thank Veronica Akle and Yefferson Ardila for the zebrafish samples.

## Supporting information captions

Fig S1.1: Collimation of the LED.

Fig S1.2: Placement of the field diaphragm at the focal plane of the collector lens.

Fig S1.3: Placement of the first spherical lens.

Fig S1.4: Placement of the aperture diaphragm at the back focal plane of the condenser lens.

Fig S1.5: Placement of the first spherical lens.

Fig S1.6: Coupling the illumination to the rest of the microscope.

Supplementary table 1: Low cost illuminator and condenser using a self-made LED illuminator to cut costs significantly as well as lower cost parts.

Supplementary table 2: Optional parts for the Illuminator for reduced spaces where a straight illuminator cannot be used and a translation stage to facilitate focussing of the filed diaphragm.

Supplementary table 3: Cost of commercial condenser and illuminator

Fig S3.1**. Contrast and depth of field vs NA.** By closing the aperture iris we can control the numerical aperture of the illumination. A piece of thin paper tissue as seen through a 10x NA 0.3 air objective A) with the aperture iris almost completely closed and B) completely open. A z-stack of the tissue was acquired, comprising 16 photographs at steps of 5um. C) and D) are the respective orthographic projections along the dotted lines in A) and B) respectively. Scale bars 50um.

Fig. S3.2. **Biological application:** Notochord of the zebrafish under different illuminations. A) The 4f system with full NA. B) A LED flashlight. C), D), regions within the dashed rectangle in A) and B), respectively. Skeletal muscle fibers are distinguishable in C).

# Supplementary material for *"Simple and open 4f Koehler transmitted illumination system for low-cost microscopic imaging and teaching"*

Jorge Madrid-Wolff[1], Manu Forero-Shelton[2]

1- Department of Biomedical Engineering, Universidad de los Andes, Bogota, Colombia
2- Department of Physics, Universidad de los Andes, Bogota, Colombia

## 1. Assembly Instructions for the transillumination condenser and light source

1) Collimate the LED light source.

Place a diffusive aspheric lens (Thorlabs, ACL2520U-DG6-A) in front of a mounted LED (Thorlabs, MWWHL4). To do so, mount the aspheric collector lens in an adjustable length lens tube (Thorlabs, SM1V05); hold it in position with a retaining ring. Attach a lens tube (Thorlabs, SM1L03) to the LED mount. Adjust the distance between the collector lens and the LED by rotating the mounts with respect to the other until the least-diverging, non-converging beam is achieved, as in figure S1.1. Fix this distance with a pair of thread stops (Thorlabs, SM1RR). Since the LED is not a point-like light source, it is not possible to completely collimate the beam. Mount the LED on a cage plate (Thorlabs, CP12) and insert cage rods (Thorlabs, ER4) to this plate.

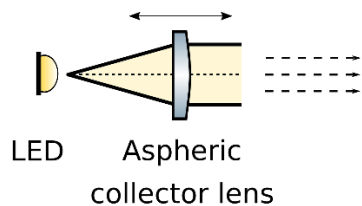

Fig S1.1: Collimation of the LED.

2) Insert the field diaphragm.

Insert a cage iris (Thorlabs, CP20S) in the cage rods. Close the iris to its minimal aperture. With the aid of a photodiode power sensor (Thorlabs, S121C) placed on the other side of the aperture, displace the diaphragm to the position of maximum intensity, as in figure S1.2. Fix the iris in position.

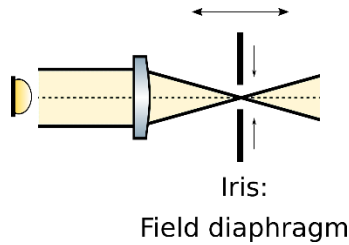

Iris:
Field diaphragm

Fig S1.2: Placement of the field diaphragm at the focal plane of the collector lens.

3) Insert the first lens.

Place the first lens (Edmund Optics, f=50mm achromat) within a cage plate (Thorlabs, CP02/M) and insert the latter in the cage rods. While keeping the field diaphragm closed, as in figure S1.3, adjust the distance from the achromat to the iris until a collimated beam is achieved. Fix the position of the lens.

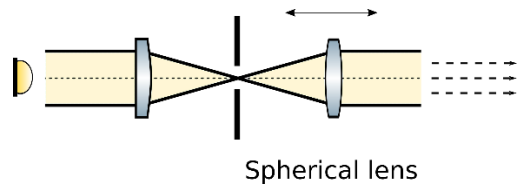

Spherical lens

Fig S1.3: Placement of the first spherical lens.

4) Insert the aperture diaphragm.

Open the field diaphragm to its full aperture and insert a cage iris (Thorlabs, CP20S) after the first spherical lens. Close the second iris (the aperture diaphragm) to its minimal aperture. Analogously to step 2, adjust the position of the aperture diaphragm until achieving the maximal intensity on the photodiode power sensor. Fix the iris in position, as in figure S1.4.

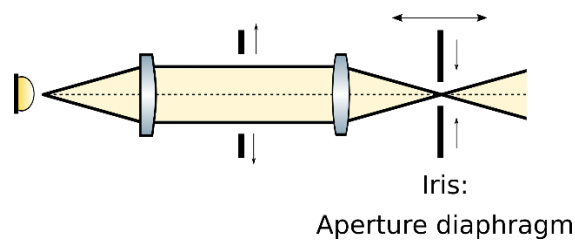

Iris:
Aperture diaphragm

Fig S1.4: Placement of the aperture diaphragm at the back focal plane of the condenser lens.

5) Insert the condenser lens.

Keep the field diaphragm open and the aperture diaphragm closed and insert the second lens in the condenser (Edmund Optics, f=40mm achromat) in the cage rods. Like step 3, adjust the lens' position until a collimated beam is achieved, as in figure S1.5. Fix the position of the lens.

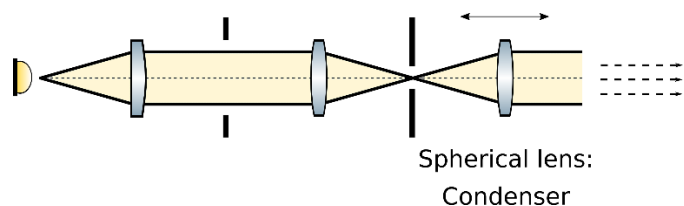

Fig S1.5: Placement of the first spherical lens.

6) Align the illumination system with the rest of the microscope.

Mount the system on a 12.7mm optical post (Thorlabs, TR50/m) and 12.7mm post holder (Thorlabs, PH50/M). Close the field diaphragm to its minimal aperture. Displace the whole system sideways and along the optical axis until an image of the field diaphragm appears in the microscope's camera or eyepiece, as in figure S1.6. Fix the system to this position.

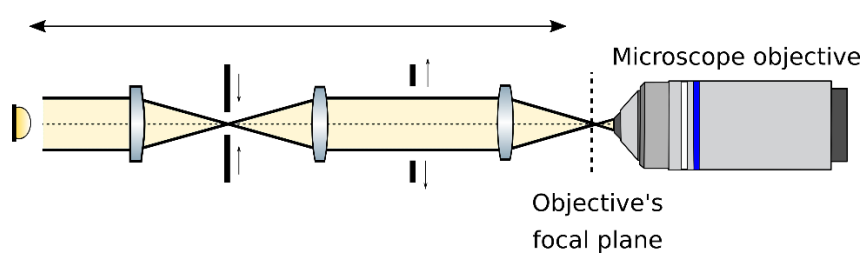

Fig S1.6: Coupling the illumination to the rest of the microscope.

7) Fine alignment of the illumination LED (optional)

With the system in place, use a digital camera in live view and get a profile of the intensity across the field. This live profile can be obtained using µManager (1) and ImageJ (2) jointly, by choosing live view, drawing a line across the field of view, plotting the intensity profile and setting it to *live* mode. Adjust the position of the LED diode as in step 1 by rotating the lens tube (Thorlabs, SM1L03) until you achieve the flattest intensity profile and then lock the lens tube in place with the outer rings.

## 2. Cost of the system

### Supplementary table 1: Low-cost illuminator and condenser

| Supplier | Description | Part Number | Quantity | Price (USD) | Sub Total | Website | Notes |
|---|---|---|---|---|---|---|---|
| | **LED illumination** | | | | | | |
| mactronica | Keyes LED 3W arduino module | generic | 1 | 3 | 3 | https://www.mactronica.com.o | |
| mactronica | DC-DC converter | LM2596 | 1 | 2 | 2 | https://www.mactronica.com.o | |
| Sigmaelectro | LM358 op amp | LM358 | 2 | 0.3 | 0.6 | https://www.sigmaelectronica. | |
| mactronica | Power supply for LED controller 12V 1A | WFI-1210 | 1 | 3 | 3 | https://www.mactronica.com.o | |
| mactronica | Cooling fins from a 3D printer head | F3D V6 | 1 | 6 | 6 | https://www.mactronica.com.o | |
| mactronica | Cooling fan for fins | generic | 1 | 3 | 3 | https://www.mactronica.com.o | |
| mactronica | Other electronic parts | generic | 1 | 3 | 3 | | cables, potentiometer, board, etc.. See note for details |
| | 3D printed parts | see drawings | 3 | 2 | 6 | | |
| Thorlabs | Adjustable Lens Tube | SM1V05 | 1 | 30.25 | 30.25 | duct.cfm?partnumber=SM1V05 | |
| Thorlabs | Aspheric Condenser Lens w/ Diffuser, Ø2 | ACL2520U-DG6 | 1 | 18.46 | 18.46 | m?partnumber=ACL2520U-DG6 | Light collector. Make sure the round f |
| | **Koehler Optics** | | | | | | |
| Thorlabs | Cage plate (lens mount) | CP02/M | 4 | 16.4 | 65.6 | duct.cfm?partnumber=CP02/M | Mount the koehler lenses and irises |
| Thorlabs | Cage plate (lens mount) Thick | CP02T/M | 1 | 21.93 | 21.93 | uct.cfm?partnumber=CP02T/M | The support holds on to this plate |
| Thorlabs | SM1 Lever-Actuated Iris Diaphragm | SM1D12 | 2 | 58.14 | 116.28 | duct.cfm?partnumber=SM1D12 | |
| Thorlabs | cage rods 18" | ER18 | 3 | 25.5 | 76.5 | product.cfm?partnumber=ER18 | |
| Thorlabs | F=50 Lens | LA1131 | 2 | 21.6 | 43.2 | duct.cfm?partnumber=LA1131 | |
| | **Support** | | | | | | |
| Thorlabs | Optical post | TR50/M | 1 | 5.19 | 5.19 | duct.cfm?partnumber=TR50/M | |
| Thorlabs | Optical post holder | PH40E/M | 1 | 23.97 | 23.97 | number=PH40E/M#ad-image-0 | These are magnetic post holders, that make alignment relatively easy. Alternatively you can buy nonmagnetic ones and forks |
| | | | | Total | 427.98 | | |

Supplementary table 1: Low cost illuminator and condenser using a self-made LED illuminator to cut costs significantly as well as lower cost parts.

### Supplementary table 2: Optional parts

| Supplier | Description | Part Number | Qty | Price | Sub Total | Notes |
|---|---|---|---|---|---|---|
| | **Optional parts** | | | | | |
| | **90° rotation** | | | | | Makes the system more compact in case space is limited |
| Thorlabs | Round Protected Aluminum Mirror | MF1-G01 | 1 | 13.97 | 13.97 duct.cfm?partnumber=MF1-G01 | |
| Thorlabs | Right-angle kinematic mirror mount | KCB1/M | 1 | 143 | 143 artnumber=KCB1/M#ad-image-0 | Cheaper alternative: parts C45M1 and a CP30 for $125, makes it easier to see components |
| | **Alignment translator** | | | | | Makes alignment easier, any compatible translation can be used. These attach to an optical bench |
| Thorlabs | Translation Stage | PT1B/M | 1 | 215.2 | 215.22 artnumber=PT1B/M#ad-image-0 | Order a non-metric version if you are in the US. Same with the bolts. |
| Thorlabs | Clamping Fork | CF125 | 1 | 8.95 | 8.95 partnumber=CF125#ad-image-0 | Stage is non-magnetic so a fork is required to attach the system |
| Thorlabs | M6 x 1.0 Stainless Steel Cap Screw, 12 mm Long, pack of 25 | SH6MS12 | 1 | 8.11 | 8.11 duct.cfm?partnumber=SH6MS12 | Only 3 are needed, can be found in hardware stores |

Supplementary table 2: Optional parts for the Illuminator for reduced spaces where a straight illuminator cannot be used and a translation stage to facilitate focussing of the filed diaphragm.

### Supplementary table 3: Cost of commercial condenser and illuminator

| Supplier | Part Number | Item | Product URL | Qty | Price | Subtotal |
|---|---|---|---|---|---|---|
| Thorlabs | WFA1000 | Transmitted Light Illumination / DIC Imaging Module 30 mm Cage Compatible | https://www.thor | 1 | 4233 | 4233 |
| Thorlabs | WFA0150 | 95 mm Dovetail Clamp for WFA1000 and WFA1100 Modules | https://www.thor | 1 | 325 | 325 |
| Thorlabs | WFA1010 | Warm White Illumination Kit | https://www.thor | 1 | 825 | 825 |
| Thorlabs | LEDD1B | T-Cube LED Driver 1200 mA Max Drive Current (Power Supply Not Included) | https://www.thor | 1 | 304,98 | 304,98 |
| Thorlabs | KPS101 | 15 V 2.4 A Power Supply Unit with 3.5 mm Jack Connector for One K- or T-Cube | https://www.thor | 1 | 33,33 | 33,33 |
| Thorlabs | CSC2001 | LWD Condenser 0.78 NA Male D3N Dovetail 400 - 850 nm One C32 Tray Included | https://www.thor | 1 | 1285 | 1285 |
| | | | | | Total (USD) | 7006,31 |

Supplementary table 3: Cost of commercial condenser and illuminator

## 3. Imaging capabilities with the system

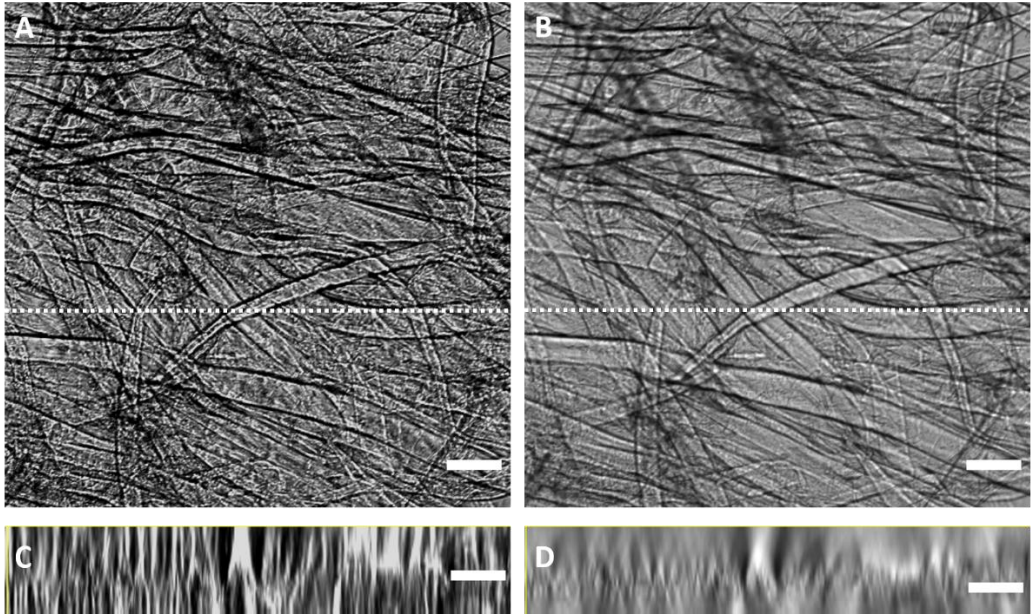

Fig S3.1. **Contrast and depth of field vs NA.** By closing the aperture iris we can control the numerical aperture of the illumination. A piece of thin paper tissue as seen through a 10x NA 0.3 air objective A) with the aperture iris almost completely closed and B) completely open. A z-stack of the tissue was acquired, comprising 16 photographs at steps of 5um. C) and D) are the respective orthographic projections along the dotted lines in A) and B) respectively. Scale bars 50um.

Figure S3.1 shows the effect of adjusting the numerical aperture of the illumination. Reducing the aperture increases contrast, which facilitates to distinguish one feature from another laterally, but lengthens depth of field, thus reducing the ability to distinguish between different planes. Note that in figures S3.1C and S3.1D, which are the orthographic projections of a *z*-stack along the dotted lines in S3.1A and S3.1B, cones of confusion appear more elongated in S3.1C than in S3.1D. This means that one can more accurately tell the depth of a given feature when high NA illumination is used.

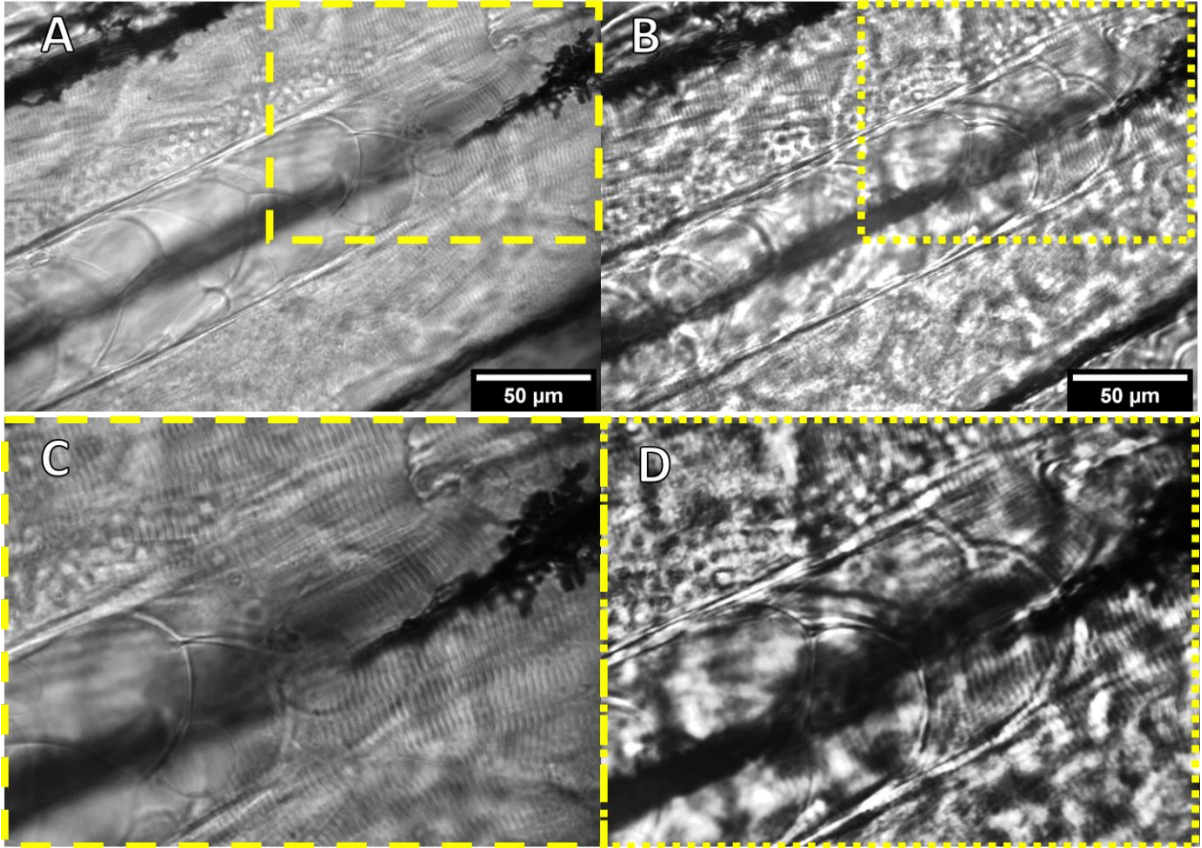

Fig. S3.2. **Biological application:** Notochord of the zebrafish under different illuminations. A) The 4f system with full NA. B) A LED flashlight. C), D), regions within the dashed rectangle in A) and B), respectively. Skeletal muscle fibers are distinguishable in C).